%% file: main.tex
\documentclass[12pt]{article}
\usepackage{authblk}
\usepackage[utf8]{inputenc}
\usepackage{amsmath}
\usepackage{url}
\usepackage{enumerate}
\usepackage{graphicx}
\usepackage{color}
\usepackage{paralist}
\usepackage{hyperref}
\usepackage{lscape}
\usepackage{rotating}
\usepackage[shortlabels]{enumitem}
\usepackage{scalerel}
\usepackage{natbib}
\usepackage{soul}
\usepackage{amssymb}
\DeclareMathOperator{\bigtimes}{\scalerel*{\times}{\sum}}
\input{bknot}

\title{A comparison of graphical methods in the case of the murder of Meredith Kercher}
\author[1]{A. Philip Dawid}
\author[2]{Francesco Dotto}
\author[3]{Maxine Graves}
\author[3]{Joseph B. Kadane}
\author[4]{Julia Mortera}
\author[5]{Gail Robertson}
\author[6]{Jim Q. Smith}
\author[7]{Amy L. Wilson}
\affil[1]{University of Cambridge, UK}
\affil[2]{Universit\`a Roma Tre, Italy}
\affil[3]{ Carnegie Mellon University, USA}
\affil[4]{ University of Bristol, UK}
\affil[5]{  Biomathematics \& Statistics, Scotland}
\affil[6]{ University of Warwick, UK}
\affil[7]{ University of Edinburgh, UK}

\date{\today}

\begin{document}
	
	\maketitle
	\begin{abstract}
	 We compare three graphical methods for displaying evidence in a legal case: Wigmore Charts, Bayesian  Networks and Chain Event Graphs. We find that these methods are aimed at three distinct audiences, respectively lawyers, forensic scientists and the police. The methods are illustrated using part of the evidence in the case of the murder of Meredith Kercher. More specifically, we focus on representing  the list  of propositions, evidence, testimony and facts given in the  first trial against Raffaele Sollecito and Amanda Knox  with these graphical methodologies. 
	\end{abstract}
	
\noindent {\small {\em Some key words:}  Amanda Knox, Bayesian networks, Chain event graphs, Meredith Kercher's murder case, Raffaele Sollecito,  Wigmore Charts}

	\section{Introduction}

	Here we explore the usefulness of three graphical methodologies, namely Wigmore Charts (WCs), Bayes Networks (BNs), and Chain Event Graphs (CEGs), in understanding and representing  the intricacies of judicial proceedings. We illustrate these methodologies on the first trial for  Meredith Kercher's murder. More specifically, we focus on representing  the list  of propositions, evidence, testimony and facts given in the trial (\citealt{sentence}) with these graphical methodologies. In this first court proceeding Sollecito and Knox were found guilty and sentenced to 25 and 26 years, respectively. We wish to clarify from the start of this paper, that we  use the proceedings of the first trial only to illustrate the different graphical models. We are not expressing  opinions about the merits of the arguments put forward.

	

	Graphical representations are powerful tools that can be applied to a multitude of settings, including legal processes. 
	Below is a list that details and supports this claim
	\begin{itemize}
		\item They provide a framework for eliciting and then representing pertinent information about a case using a well-defined and unambiguous semantic that converts an (often large) collection of natural language statements into graphs that gives a wholistic picture of a case.
		\item They provide a framework which can be embellished into a model that gives a full description of a case.
		\item They provide a framework around which, with suitable embellishments, various useful explanations can be synthesized -- for example, the most compelling explanation why someone is guilty or innocent.
		\item They give a framework for describing to third parties why such deductions are true in a way that is understandable and transparent to someone who uses the graphical model.
	\end{itemize}
	
	Given its generality, this list applies to the three graphical methodologies we illustrate here. These, however, differ in what they represent. Broadly speaking: Wigmore charts graph arguments; Bayesian networks depict items evidence, propositions and the relationship among them;  Chain event graphs depict time dependent actions and stories. In turn, this makes each graph especially useful to specific audiences, \textit{i.e.} lawyers, forensic scientists and the police, respectively. Here we compare the utility of the three graphical methods for assessing the propositions associated with a real case, in which Amanda Knox and Raeffele Sollecito were accused of the murder of Meredith Kercher in Perugia, Italy in 2007. 
 

	The paper is organised as follows. In Section 2, we present a  review of the graphical methods. Section 3 gives a summary of the Meredith Kercher case. Sections 4, 5 and 6 are devoted to representing the Kercher case using Wigmore Charts,  Bayesian Networks and  Chain Event Graphs, respectively.  Section 7 compares the three graphical methods and a discussion on the highlights and  limitations of each method is given in Section 8. 
The work in this paper was divided into different teams, Kadane and Graves worked on WCs, Dawid, Dotto and Mortera on BNs and Robertson, Smith and Wilson on CEGs. 
	\section{Graphical methods for assessing propositions}
	\label{sec:graphs}
	We now give a brief  overview  of the three different graphical methods we will use for the case: WCs, BNs and CEGs.

	\subsection{Wigmore Charts}
	A Wigmore Chart is a way of creating a visual representation of how legal arguments relate to each other and to the questions before a court. Facts may be presented in court in the form X says Y about Z. A Wigmore Chart encodes answers to questions such as :
	\begin{itemize}
		\item Is X in general truthful ?
		
		\item Was X in a position to know Y about Z ?
		
		\item Is there contrary evidence about Y in relation to Z ?
		
		\item Is there other supporting evidence about Y  in relation to Z ?
		
		\item How might the Y-ness of Z relate to the questions before the court ?
		
		\item What inferential steps must be accepted to proceed from X's testimony to the conclusion the court is to contemplate ?
	\end{itemize}
	
	The encoding is done by drawing lines between evidence and conclusions, identifying the argumentative steps along the way.
	
	\subsection{Bayesian Networks}
	
	A Bayesian Network (BN) is a diagrammatic representation of posited
	relationships between a collection of variables.  In a legal context,
	the variables might include hypotheses of interest and items of
	evidence, as well as additional unobservable variables introduced to
	assist in the structuring.
	
	The variables are represented by nodes of the network, and their
	relationships indicated by arrows between the nodes.  The
	interpretation is that each variable can depend on those other
	variables which feed an arrow into it (its ``parent''), but not
	otherwise on earlier variables.
	
	The dependence may be interpreted in various ways, including:
	\begin{itemize}
		\item Temporally: describing what follows what.
		\item Qualitatively: describing what is relevant to what.
		\item Quantitatively: describing the deterministic or probabilistic
		dependence of a variable on its parents.
		\item Causally: describing what is caused by what.  Here ``cause'' may be 
		\begin{itemize}
			\item interpreted intuitively, or 
			\item given an explicit interpretation, e.g. whether, and how, a
			variable might change when another variable is intervened upon.
		\end{itemize}
	\end{itemize}
	A BN has a formal semantics which allows one to deduce indirect
	implications of the assumptions built into it.  In particular, it
	supports inference about the impact of observed evidence on
	hypotheses.

\input{BNLongOverview}
	\subsubsection{Object-Oriented Bayesian Networks}
	\input{OOBNOverview}

	\subsection{Chain Event Graphs}
		\label{sec:CEG}
	\input{CEG_sectionla}

	\section{The Meredith Kercher case}
	\label{MKcase}
	
	\subsection{Background}
\label{sec:back}

In 2007 Meredith Kercher, an  Erasmus student from England, shared a flat with Amanda Knox and two other Italian flatmates in Perugia, Italy. Kercher was murdered in her bedroom on the night between the 1\textsuperscript{st} and the 2\textsuperscript{nd} of November 2007. Her body was found the day after by the police who were called by Amanda Knox and Raffaele Sollecito, her Italian boyfriend. 
They both claimed that they had spent the night at Sollecito's apartment and that when they had returned to Knox's apartment, they had found a broken window and they suspected that there had been a theft. The police found Kerchers' body inside of the apartment lying on the floor covered by a duvet, but no murder weapon was found. On 6\textsuperscript{th} of  November the police arrested Knox, Sollecito and Patrick Lumumba, Knox's employer, whom Knox, in the police interrogation accused of the murder. A few days later, on the 20\textsuperscript{th} of November, Lumumba was released after investigations revealed that he had an alibi. On the same day, the police arrested Rudy Guede, an Ivorian who had met Kercher a few days before the murder. He was suspected of the murder as his fingerprints were found at the crime scene. Guede admitted his presence at the apartment during the night of the murder but has always claimed his innocence. He was tried separately as he opted for a shortened proceeding. He was sentenced, in October 2008,  to 30 years imprisonment  for sexual assault and conspiracy to commit murder. Later his  sentence was reduced by the court of appeal to 16 years of imprisonment.

Knox and Sollecito's trial began on the 18\textsuperscript{th}  of January 2009.  In December 2009, the two defendants were sentenced, respectively, to 26 and 25 years imprisonment for the murder of Meredith Kercher. Roughly two years later, in October 2011, the Assize Court of Appeal acquitted the two defendants of the murder ``for not having committed the crime'' and ordered their release from prison due to the unreliability of the DNA traces on the assumed murder weapon. On the 25\textsuperscript{th}  of March 2013 the Supreme Court annulled the  sentence on the second appeal and referred it to the Court of Appeal of Florence for a new trial which started in September 2013. Knox and Sollecito were, once again, found guilty and were sentenced, respectively, to 28 and 25 years imprisonment. Finally, on March 2015 the Cassation Court (Supreme Court) absolved Knox and Sollecito. In this paper we only consider the evidence from  the first trial of Sollecito and Knox.

There are two key items of forensic evidence linking Knox and Sollecito to the murder: Sollecito's DNA found on Kercher's bra clasp found in her bedroom 46 days after her murder and Knox's DNA found on the handle of a kitchen knife purported to be used for the murder. The knife was found in a cutlery drawer in Sollecito's apartment. Low copy DNA was found on the blade of the kitchen knife and declared to be that of Kercher. Also, Sollecito's DNA found on Kercher's bra clasp was low copy number.  It is unclear whether the police followed anti-contamination procedures when handling the clasp and the knife. For example, video footage suggests that police wore non-sterile gloves when handling the bra clasp. No DNA evidence from Knox or Sollecito was found in the bedroom or on Meredith's body, except for that of Sollecito on the bra clasp.
As in subsequent trials and discussed in \cite{GillKercher}  the DNA evidence of the knife and bra clip are very likely due to laboratory contamination.
Here we  decided not to study the DNA evidence, to exclude a major topic of contention (see \citealt{gill2014misleading, GillKercher} )

In this paper we  investigate and compare how the three graphical methods described in Section \ref{sec:graphs} can be used to assess a list of propositions associated with the Meredith Kercher case. We  do this by constructing three graphical models of the evidence and arguments that were presented as part of the first trial. To keep the focus on the comparison of the graphical methods we will restrict consideration to the arguments associated with the knife and specifically to the question of whether the knife found in Sollecito's kitchen drawer is consistent with Meredith Kercher's wounds. In the graphical representations we do not consider  the DNA evidence associated with the knife, 
	since the results of the analysis of the DNA were  deemed unreliable by the experts of the Assize Court of Appeal of Perugia who acquitted Knox and Sollecito.

The reasons for choosing this case were:
\begin{itemize}
	\item the case was very well known and many of the propositions transparent to a general reader.
	\item There is a lot of information about the case in the public domain.
	\item It is sufficiently complex to be able to illustrate many of the points we would like to make. 
\end{itemize}

Of course, these advantages also brought with them certain challenges. Much of the vast source information was in Italian and needed translating and there were many different trials. However, because this paper is not attempting to provide a balanced appraisal of all the evidence in this case, we found it sufficient to demonstrate and compare the efficacy of different types of graphically based analyses that might be available. Here, we have focused our analyses only on evidence related to the knife purportedly used as the murder weapon. 	
	
	\subsection{Evidence, testimony, propositions  and arguments modelled}
	\label{sec:thelist}

	The main item of evidence we consider is  Sollecito's knife, termed  \textit{exhibit 36} in \cite{sentence}. This knife had a 17.5 cm blade length,  was 3 cm wide and 1.5 mm thick with striations at 2.2 and 11.4 cm from the knife's tip.

	The wounds found on Kercher's body were:

	\begin{itemize}
		\item A fatal wound on the left of her neck that was 8 cm deep, 8 cm long and 0.4 cm wide.
		\item A smaller wound on the right of her neck that was  4 cm deep, 1.5 cm long and 0.4 cm wide.
	\end{itemize}
	
In particular, we use the three graphical methods to examine whether Kercher's wounds were consistent with the  use of Sollecito's knife and whether it could have been the murder weapon. 
	
The expert witnesses called in the trial for each side were:
	\begin{itemize}
		\item 	Leviero, Bacci, Lalli, Novelli, Toricelli nominated by the PM (the public prosecutor) and/or the  civil party; 
		
		\item	Introna, Torre, Patumi (for Knox) Tagliabracci (for Sollecito) nominated by the defence;
		
		\item	Others nominated by the GIP (the preliminary investigation judge).
	\end{itemize}

We now give a list of the testimony, propositions and evidence items extracted from the document of the first trial against Amanda Knox and Raffaele Sollecito for the murder of Meredith Kercher, Court of Assizes of Perugia, 4-5/12/2009 (\citealt{sentence}). The page numbers in brackets refer to  the numbers on the bottom left of the pages in the original Italian document mentioned above.
	

	\begin{enumerate}
		\item	Knox recognised the knife (exhibit 36)
		\item	Knox's admission to 1 (in interrogation of 12-13 June 2009) (63)
		\item	Knox had used the knife in Sollecito's house for cooking in his kitchen. (63)
		\item	Knox statement as to item 3. (63)
		\item	Knox had never carried the knife elsewhere. (63)
		\item	Knox statement as to item 5. (63)
		\item	There was blood on a knife found at Sollecito's house.
		\item	An inspector told Knox about item 7. in prison.
		\item	Knox was worried about item 7.
		\item	Knox's statement as to items 8. and 9. in a prison conversation between Knox and her mother (66).
		\item	Knife in Sollecito's kitchen drawer appeared to be very clean and was put in a clean envelope.
		\item	Police statement as to item 11. (99)
		\item	Cause of death: from a non-serrated knife wound, together with strangulation, suffocation and haemorrhagic shock. (106)
		\item The hyoid bone was fractured. 
		\item	Item 14. was compatible both with strangulation and with penetration by a knife.
		\item	Lalli autopsy report as to item 14. (113)
		\item	 Liviero gave  a 50\% chance that item 14. could have been caused by one or two people (113).
		\item	\textit{Exhibit 36} is compatible with the major wound.
		\item	Lalli's testimony to item 18. (113)
		\item	The knife had striations on the blade 
		\item	Liviero's testimony to item 20. (113)
		\item	Liviero could not state whether one or more persons committed crime. (113)
		\item	Bacci's testimony to item 18. (116)
		\item	Second wound on right side of neck incompatible with knife as this wound was 1.5 cm long and \textit{Exhibit 36} is at least 3 cm wide.	
		\item
		\begin{enumerate}[a]
			\item	Bacci's testimony to item 24. (116). 
			\item	Torre's testimony to item 24. (142)
		\end{enumerate}
		\item	Major left side neck wound, 8 cm long, can be made by a 3 cm wide knife (by rotation)
		\item	Norelli's testimony to item 26. (121)
		\item	Norelli's testimony to item 24. (122)
		\item	Whole length of knife entered major wound.
		\item	There was bruising at the major wound.
		\begin{enumerate}[a]
			\item	Liviero's denial of  item 30. (113)
		\end{enumerate}
		\item	The bruising was caused by the knife handle.
		\item	Introna's (defence) assertion of items 29., 30. and 31. (133/136)
		\item	Introna's denial of 18, on account of items 29. and 31.
		\item	Knife penetrated at least 2-3 times in major wound.
		\item	Torre's (defence) testimony as to item  34. (141)
		\item	A 17 cm knife would have gone right through the victim's neck and not made only an 8cm wound.
		\item
		\begin{enumerate}[a]
			\item	Introna (defence) assertion of 36 and consequent denial of 18. (132)
			\item	Torre (defence) assertion of 36 and consequent denial of 18. (142)
		\end{enumerate}
		\item	Traces of blood on mattress cover, made by a knife of maximum width 1.4cm.
		\item	Vinci's testimony to item 38. (146)
		\item	A single knife was used. 
		\item	Vinci's testimony of compatibility with item  40. (146)
		\item	Ronchi's testimony of compatibility with item 40 (and with \textit{exhibit 36}). (148)
		\item	Cingolani's testimony to item  24. (151)  
		\item	Cingolani testimony to item  18. (153)
		\item	Cingolani not sure about item  31.  (153)
		\item	Cingolani could not see striations (cf item 20.). (153)
		\item	Patumi's testimony to item  24. (156) 
		\item	Patumi's testimony to item  30. (156)
		\item	Patumi considers item  31. probable. (156)
	\end{enumerate}
	
	\section{Wigmore Chart}
 \label{sec:wigmore}

	 A Wigmore chart\footnote{Wigmore charts were invented by John H. Wigmore, dean of the Northwestern University Law School.} sketches how items of evidence relate to the propositions that must be proved (probanda) in a case. Wigmore's original version is very complicated with many different symbols reflecting categorizations of the items graphed. He put probanda at the top of the page, and used vertical lines to represent arguments leading to the next step, and horizontal lines to indicate contradictory arguments or evidence. That convention is also used in 
	\cite{kadane2011probabilistic}
	in an analysis of the Sacco and Vanzetti case. See also \citet{anderson1998analysis}. 

	We are using the evidence concerning Sollecito's knife as an example of the use of Wigmore charts. There are several limitations in our doing so:
	
	\begin{enumerate}
		\item There were five court decisions related to the murder charges against Sollecito and Knox with respect to the murder of Meredith Kercher (see Section \ref{sec:back}). We are using only the record of the first of these.
		\item We are dealing only with the court's view of the evidence and arguments. Specifically we do not have a transcript which records, word-for-word what was said in court. So we do not have the prosecution and defense summaries of the evidence, that would show, most starkly, the chain of reasoning each is proposing.
	\end{enumerate}
	
	 Evidence not relevant to probanda, and hence not leading to or away from a probandum, are irrelevant to the case.
	The first thing to decide is the probandum to which the listed evidence mainly relates. There was no evidence listed that directly related to the probandum  “Sollecito's knife was used to kill Kercher”. Thus,  the narrower sub-probandum  “The use of Sollecito's knife is consistent with Kercher's wounds” was used. The difference between these probanda is that, if the sub-probandum were established or believed, there would still be many other knives also consistent with Kercher's wounds.
	
		Because Wigmore Charts are about arguments arising from evidence, they can require items to be added that are not evidence \textit{per se}, but rather steps in a plausible chain of reasoning leading from one or more pieces of evidence to a probandum.
		
	One of the major virtues of Wigmore charts is that they make evident items of evidence that do not bear on the probandum. With this list of items of evidence and this probandum, there are several of  these. Chart 1 displays these.

	\begin{figure}
		\centering
		\includegraphics[width=6in]{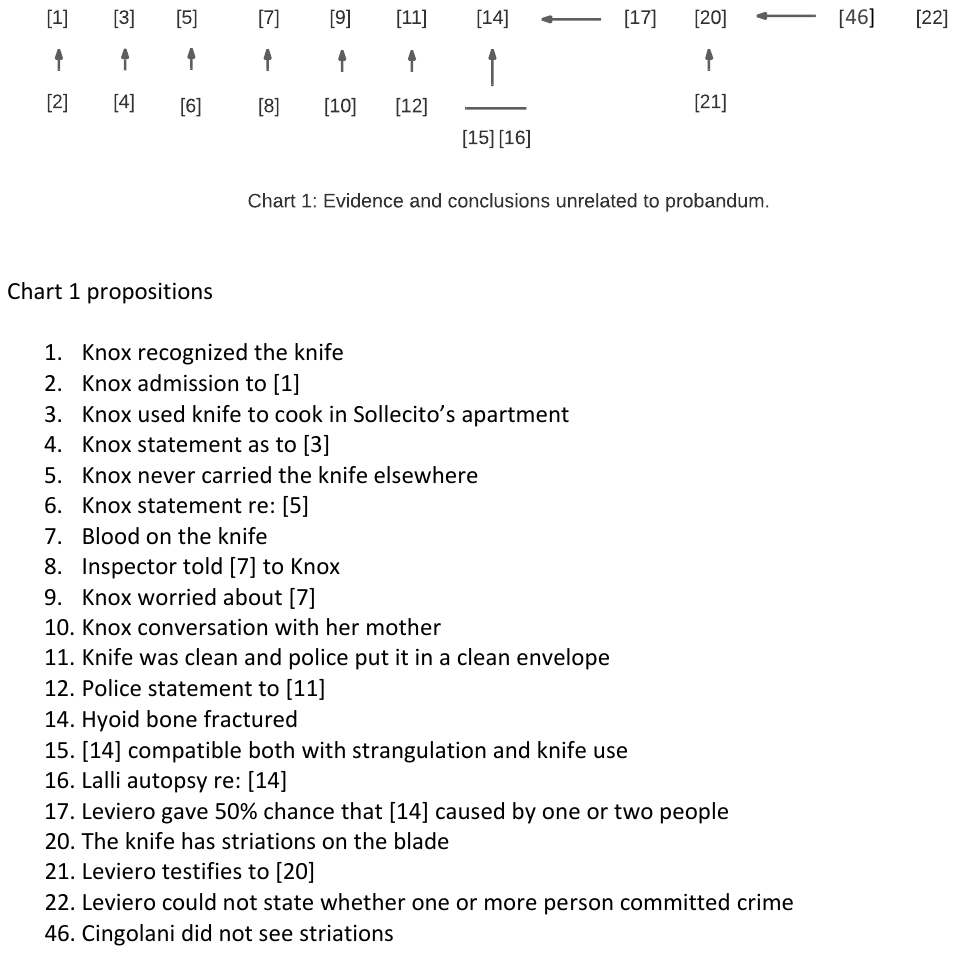}
	\end{figure}

	Kercher's death, according to the autopsy, \textit{item 13} in Section \ref{sec:thelist},  was due to a wound from a non-serrated knife, together with strangulation, suffocation and hemorrhagic shock. She had two wounds, a major one and a minor one. Chart 2 gives the evidence for and against the sub-probandum “Solliecito's knife is consistent with Kercher's major wound.”

	\begin{figure}
		\centering
		\includegraphics[width=6in]{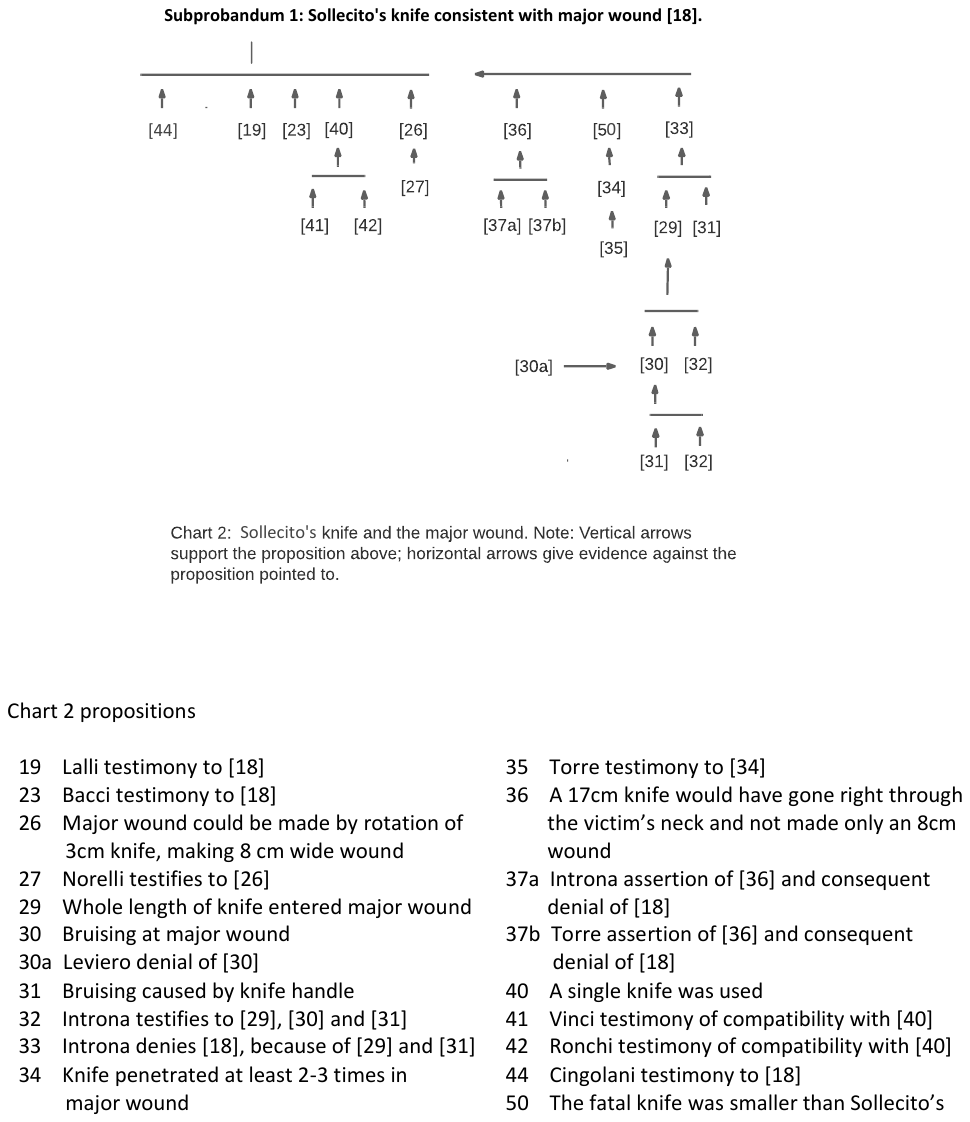}
	\end{figure}

	\begin{figure}
		\centering
		\includegraphics[width=6in]{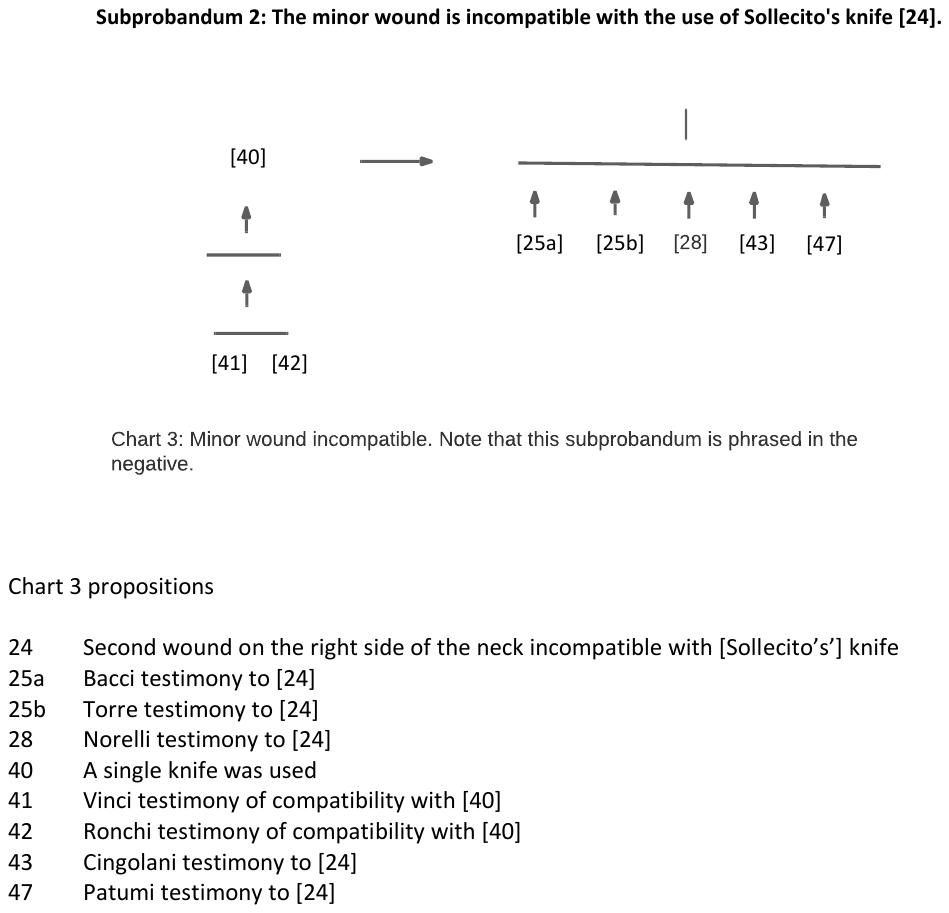}
		\label{fig:3}
	\end{figure}

	One of the principal virtues of Wigmore charts is that they can be used to make explicit unspoken assumptions in a chain of reasoning. (For several examples, see \citet{kadane2011probabilistic}). Most of the items of evidence listed in this case are not of that character, in that they do not reveal a chain of reasoning. However, there is one exception: the claim, shown in Chart 2, that the knife penetrated the major wound 2 or 3 times. What has this to do with the Subprobandum~1: ''Solliecito's knife is consistent with the major wound''?  For this reason an additional item 51. which states that the fatal knife is smaller than Solliecito's knife, was added to Chart 2. This is thus an argument against Subprobandum~1. Most of the items in the list of fifty are conclusory, and do not display the reasoning behind the conclusion.   Hence, they are not very interesting for display in a Wigmore chart. 
	
While Chart 2 addresses evidence relevant to whether Sollicito's knife is consistent with Kercher's major wound, Chart 3 addresses the subprobandum that her minor wound is incompatible with the use of Sollecito's knife.
	
	\section{Bayesian Network}
	\input{BN}

	\section{Chain Event Graph}

\input{CEG}

	\section{Comparison of the three methods}

 There are many similarities between a Bayes Network and a Chain Event Graph in that both are graphical representations of a probability model, unlike a Wigmore Chart which is generally a graphical representation of a set of legal arguments (see further discussion on this in Section \ref{sec:discussion}). Thus, both BNs and CEGs can be used to estimate probabilities and other statistics (such as the likelihood ratio) quantitatively, whereas a WC provides a more qualitative description of the propositions and arguments in a case.  Both BNs and CEGs do not only model probabilistic dependence, they can also model logical dependence.
 
 One important difference between a BN and a CEG is that a CEG has the ability to represent temporal information - this is because events can be represented as unfolding edges through time. CEGs also have an advantage when modelling asymmetric details in a case (such as dead ends), in that the chains describing a particular unfolding can just stop. In a BN it would be necessary to continue enumerating situations that did not occur in the conditional distributions associated with the nodes. One downside of this approach however is that CEGs are generally less compact than BNs due to this very representation of lots of different propositions and possibilities with individual edges (this is particularly a problem when drawing the staged tree as a precursor to the CEG). As such there is a need for software packages to help draw and perform computations with CEGs. For BNs many such packages already exist.

All three of the graphical approaches aim to document a mental model of how different aspects of a case relate to each other. People associated with a case can then use the networks to work through and understand different possible arguments and how the available evidence supports the various propositions. One important point is that for all three approaches, the mental model is that of the person who drew the network. Another person would likely create a different version. This is illustrated by the fact that all three of the example networks for the Kercher case make slightly different arguments and draw different links between the statements in Section \ref{sec:thelist}.

For both the BN and the CEG being able to understand and represent the conditional probabilistic structure in the case may be the biggest challenge for practitioners looking to use these approaches. For WCs, specifying this structure is not necessary.  Making probabilistic arguments in a WC can be possible, however it cannot be used as an inference engine as with a BN and a CEG. There is a question of whether the conditional dependencies encoded in the networks are fully understood by users and hence whether they are completely transparent. As a result, people with different backgrounds might find different graphical approaches useful. Those with knowledge of probability might prefer a BN or CEG and lawyers (used to legal reasoning) might prefer a WC. 

The propositions concerning Sollecito’s knife in the BNs and CEGs were chosen to represent the arguments made in the first trial at the time, whether true or fallacious. As a point of  fact there was very little evidence that could justifiably link Sollecito’s knife to the wounds. The arguments associated with the size of the wounds can only speak to whether or not a knife with the same dimensions as Sollecito’s knife could have inflicted the wounds.

 The blade of Sollecito's knife was examined for chemical residues, and starch was found. A police inspector lied to Knox (who was jailed) that they had found blood on the knife. In a telephone call, Knox told her mother that this worried her. The fact that Knox was worried by this report was regarded by the Court as evidence for the possible use of Sollecito's knife as the murder weapon. As a result, it was necessary to include
 in the initial propositions the possibility that Sollecito’s knife itself was used in the attack. The possibility that an alternative knife with the same dimensions as Sollecito’s knife was used has also been included in the BN and CEG. This issue with the propositions demonstrates the value of using graphical approaches to set out the reasoning in a case – by thinking through the links between evidence and propositions and the associated probabilities it is clear which pieces of evidence are relevant to the propositions being considered.  

		
	
	A difference between a BN and a WC 
	is that an arrow in a WC follows the direction   of a desired inference, whereas in a BN it follows the direction of
 the supposed process or of the cause-effect relationship.  The latter may be regarded   as more stable, and easier to specify, than an inferential
  relationship.
Furthermore,  a fully specified BN supports probabilistic inference from
  evidence to hypothesis, even when this flows against the arrows.
	
	\section{Discussion}
	\label{sec:discussion}
	\subsection{Limitations of each graphical model }
	
	\textbf{Wigmore Charts}

	Wigmore Charts organize the evidence adduced in a trial, and how (and whether) there is a plausible chain of argument leading to a probandum. In itself, it does not address how plausible that chain is. In an analysis of the case against Nicola Sacco and Bartolomeo Vanzetti for the robbery and murder of Frederick Parmeter and Alessandro Berardelli,  \cite{kadane2011probabilistic} used Wigmore Charts to parse the evidence introduced at  trial, and also evidence found later. Subjective probabilities (expressed as odds) are given for the probative force were given, separately, by the authors and by an historian of the case.
	
	But odds may not be a natural tool to use in association with Wigmore Charts. Kadane and Schum  explored the connection as best as they could at the time, but it remains an open question whether and how to use probabalistic thinking about Wigmore Charts. In this paper,we explored a small part of the evidence concerning the murder of Meredith Kercher using only a (very) simplified version of Wigmore Charts, without introducing quantitative methods.

	\textbf{Bayesian Networks}

	The construction of a Bayesian network is an iterative ongoing
	process, and that described here is in no sense final.  There are many
	choices to be made, and different individuals or teams would almost
	certainly develop quite different networks.  Even deciding which
	variables to represent in the network involves a good degree of
	personal whim; specifying appropriate states for them, and the links
	between nodes, while seemingly a simple matter of representing the
	actual situation, in fact again require many personal and somewhat
	arbitrary choices.  In principle the network should model all the
	possible situations, before taking account of the evidence; in
	practice, knowledge of what evidence is available will tightly
	constrain the modelling process.

   Like all  probabilistic forms of reasoning, including  CEGs, BNs are sensitive to changes in assigned probabilities. 
   As there is no unique way to specify conditional probabilities for each node, this enables the BN to be used to model different opinions.   Because of the
	difficulty in specifying defensible probability values, here we have
	emphasised the purely qualitative structure of the network---even
	though this is in no sense unique---taking the view that this carries
	helpful information about the complex relationships between the
	evidence and the probandum.

	\textbf{Chain Event Graphs}

	There are two main limitations to the Chain Event Graph. First, topologically it is not as compact as a BN, especially for symmetric problems. In turn this means it can be much more difficult to read and more cumbersome. As a result, this issue comes to the fore whenever competing hypothesized developments are very different to one another in terms, for example, of the relevant covariates that different hypotheses entail and the amount of detail required to specify them. Second, the available supporting code for CEGs is much more limited than that for BNs - such developments are about 30 years behind the BN which has many commercial and open supporting software tools. 	Another drawback about a CEG is that we need to input continuous evidence indirectly. 
	
	The above being said, one big strength of the CEG is that it can explicitly express the order in which events are hypothesized to have happened. If such issues are not critical to inference then a CEG is a less powerful tool. This is why they are especially useful for activity hypotheses.

	One positive thing about the use of a CEG and BN together is that a CEG can often provide a BN with an automatic way of constructing ``natural'' variables. The vertices along the cuts of the CEG define the atoms of a random variable that distinguishes the different possibilities of developments that might have different outcomes. So the set of paths form the root into such a vertex give the set of unfoldings that label each value of this random variable. One you have these then you can give it an appropriate logical meaning and name. And so name the variable. This might seem contrived but it moves from something of an art form -- which is often needed in the construction of a BN -- to something more like a protocol/algorithm for construction of the relevant random variables. This, obviously,  only works when then underlying explanation is about what might have happened.

	\subsection{Conclusions}

	 The three graphical representations we have presented highlight different aspects of a  case,  each with its own advantages and disadvantages. 
Each representation is valuable in its own way, emphasising specific aspects of the arguments, and supplying ways of sifting and analysing the intricacies of the case.  Taken together, they can work harmoniously to provide a better overview of a case.  In the future one could hope to develop formal ways of moving between the different representations,  so that understandings gleaned from one could help structure and embellish another---there are already some relevant materials, such as in \cite{Dawid2011-DAWIN}.

We have focused on a very small part of the Kercher case,  A bald representation of the total evidence would be very large and unwieldy, but, by analogy with an object-oriented approach, it might be broken down into a number of separate modules, loosely connected together.   In particular, we have not  here considered the DNA evidence in the case: that could be structured in an independent graphical module, which could then be linked to that developed here. 

Moreover, the scope of this paper has been limited to a relatively small list of propositions from the first ruling.  In total there were five rulings.  Given the complexity and length of the court case, a decision to focus solely on the first ruling was made for the sake of time and manageability. 

Extracting information from court records is an arduous task---all the more so when, as here, these are in a language foreign to many of the authors.  The document was  (poorly) translated using Google Translate.  The Italian authors translated parts of the original document by hand and  checked  all instances where the the knife (exhibit 36)  was mentioned.   An attempt was  made to locate these sentences  using the NPL corpus manager and text analysis software  \textit{Sketch Engine} (\texttt{https://www.sketchengine.eu/}).   However, a custom-made  tool would have been invaluable, at least to generate starting points for a supervised method.

We have aimed to demonstrate the usefulness of  a variety of graphical representations for coming to an understanding of non routine cases, whether  in the investigative phase or in court.  It is to be hoped that future  technological developments may make these tools easier for  interested parties, such as lawyers, forensic scientists, police and adjudicators,
to use and communicate with.


	
\subsection*{Acknowledgements}
	This research was undertaken with the support of the research project ``Statistics and the law: Probabilistic modelling of forensic evidence'', led by Amy Wilson and  Anjali Mazumder, funded by the Alan Turing Institute under wave one of the UKRI Strategic Priorities Fund, EPSRC Grant EP/W006022/1. We also thank Barbara McGillivray for the support she gave  with Sketch Engine and Ruoyun Hui for many useful discussions on the case. 
 
\bibliographystyle{apalike}
	\bibliography{references}
\end{document}

%% file: bknot.tex
\newcommand{\HIDE}[1]{ }
\newcommand{\COMMENT}[1]{ }

\renewcommand{\eqref}[1]{\mbox{(\ref{eq:#1})}}

\newcommand{\secref}[1]{\mbox{\S$\,$\ref{sec:#1}}}

\newcommand{\figref}[1]{\mbox{Figure~\ref{fig:#1}}}

\newcommand{\algref}[1]{\mbox{Algorithm~\ref{alg:#1}}}

\newcommand{\parents}[1]{{\rm pa}(#1)}

\newcommand{\nd}{{\rm nd}}

\newcommand{\cd}{\,|\,}

\newcommand{\cip}{\mbox{$\perp\!\!\!\perp$}}

\newcommand{\ind}[3]{\mbox{$#1 \, \cip\, #2 \mid #3$}}

\newtheorem{expl}{Example}
\newtheorem{definer}{Definition}

\newtheorem{algor}{Algorithm}

\newtheorem{rem*}{Remark}

\newcommand{\halm}{\hspace*{\fill} $\Box$\par}



%% file: BNLongOverview.tex
\subsubsection{Qualitative structure}
\label{app:qual}
\begin{figure}[htbp] \centering
\resizebox{\textwidth}{!}{\includegraphics{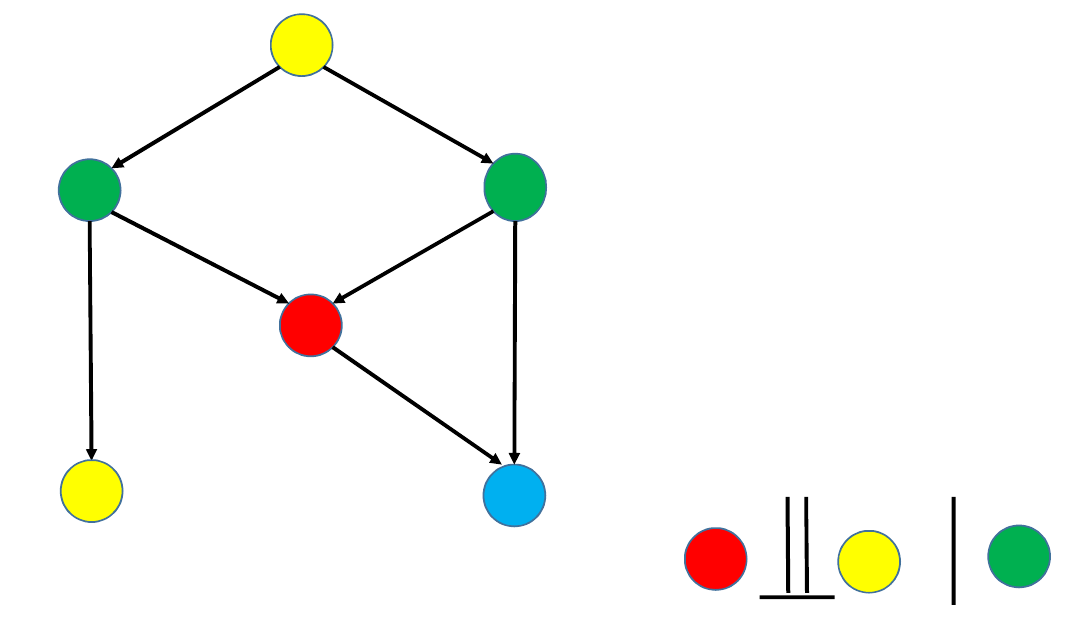}}
	\caption{A directed acyclic graph (DAG).  The red node has the
		blue node as ``child'', and the two green nodes as
		``parents''.  The red node is independent of the yellow nodes, conditional on the green nodes.}
	\label{fig:DAG}
\end{figure}

A Bayesian network (BN) is a form of {\em directed acyclic graph
	(DAG)\/}, comprising a finite set $V$ of {\em nodes\/}, with {\em
	arrows\/} between some of the nodes, in such a way that it is not
possible, by following the arrows, to return to one's starting point.
An example is shown in \figref{DAG}.
The nodes having arrows pointing out from them and into a node
$v\in V$ are called its {\em parents\/}, and denoted by
${\parents{v}}$; those with arrows pointing into them out from $v$ are
its \textit{children}.  This terminology is extended, in an obvious
way, to \textit{ancestor}, \textit{descendant}, \textit{etc.} Thus in
\figref{DAG} the red node has the blue node as its
only child, and the two green nodes as its parents.

\subsubsection{Independence properties}
\label{app:ind}
Each node $v\in V$ is identified with a random variables $X_v$, and
for $S\subseteq V$ we write $X_S$ for $(X_v: v\in S)$.  A BN {\em
	represents\/} a joint distribution $P$ over the variables
$(X_v:v\in V)$ when it is the case that, under $P$, each node in $V$
is independent of all its non-descendants, conditional on its parents we write,
in the conditional independence notation \citep{dawid1979conditional}, that
\begin{displaymath}
	\ind {X_v} {X_{\nd(v)}} {X_{\parents{v}}}.
\end{displaymath}
For example, in the distribution represented by \figref{DAG} the red
node is independent of the yellow nodes, conditional on
the green nodes (shown in the bottom right corner of \figref{DAG}).

For any such distribution $P$, we can deduce further implied
probabilistic conditional independence properties, using the following
entirely graphical routine.
\begin{algor}[Moralisation]
	\label{alg:moral}
	\begin{rm}\quad\\
		Suppose we wish to query the conditional independence property
		$\ind {X_A} {X_B} {X_C}$.  We proceed by the following steps.
		\begin{description}
			\item[Step 1: Ancestral graph] Delete from the DAG all nodes that
			are not in $A$, $B$, or $C$, or any of their ancestors.
			\item[Step 2: Moralisation] Connect by an undirected arc any
			parents of a common child that are not already connected by an
			arrow.  Then convert all arrows to undirected arcs.
			\item[Step 3: Separation] In the resulting undirected graph, look
			for a path from a node in $A$ to one in $B$ that does not enter
			$C$.
		\end{description}
		If there is no such path, deduce that, under $P$,
		$\ind {X_A} {X_B} {X_C}$.
	\end{rm}\hfill
\end{algor}

\begin{figure}[htbp] \centering
    \resizebox{\textwidth}{!}{\includegraphics{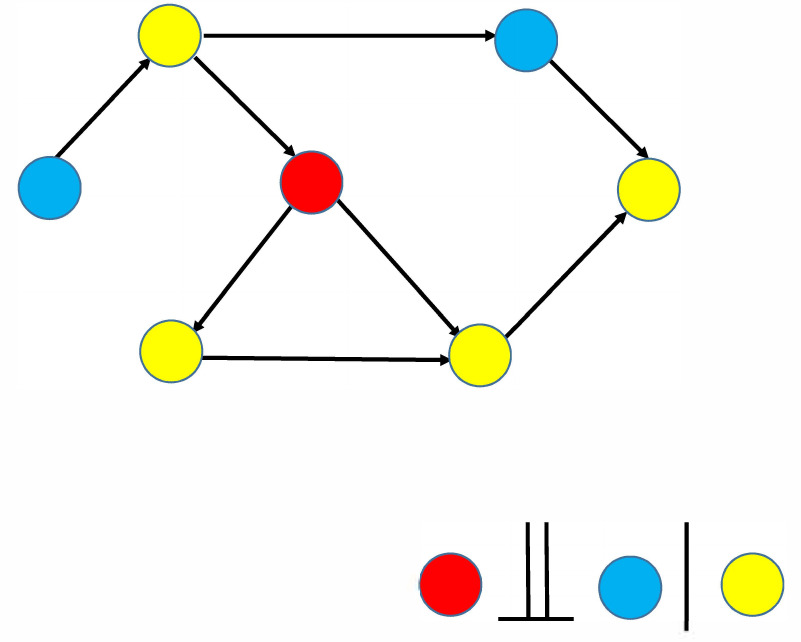}}
	\caption{Another DAG.  The red node is independent of the
		blue nodes, conditional on the yellow nodes.}
	\label{fig:DAG2}
\end{figure}

Using \algref{moral} we see that, for any distribution represented by
\figref{DAG2}, the red node is independent of the blue
nodes, conditional on the yellow nodes.

\subsubsection{Quantitative structure}
\label{app:quant}
Let ${\cal X}_v$ denote the set of {\em states\/} (possible values) of
$X_v$.  For $S\subseteq V$ we write ${\cal X}_S$ for
$\bigtimes_{i\in S}{\cal X}_i$.  For $x = (x_i:i\in V)\in {\cal X}_V$,
its projection (restriction), $(x_i:i\in S)$, to ${\cal X}_S$ is
denoted by $x_S$.

We specify, for each node $v\in V$, a distribution over ${\cal X}_v$,
conditional on each configuration
$x_{\parents{v}} \in {\cal X}_{\parents{v}}$ of the states of its
parents: the associated (discrete or continuous) density is denoted by
$p(x_v\cd x_{\parents{v}})$.  For a distribution $P$ represented by
the BN, the joint density of all variables factorises as the product
of its parent-child conditional densities:
$$
p(x) = \prod_{i\in V} p(x_i|x_{\text{pa}(i)}).
$$
Conversely any distribution with a density of this form can be
represented by the BN.

When the variables are discrete, the parent-child conditional
probability distributions can be represented as tables of the
conditional probabilities $p(x_v|x_{\text{pa}(v)})$.  The overall size
of each table is the product of the sizes of the state-spaces of the
variables concerned.

\subsubsection{Computation}
\label{sec:comp}
The structure of a BN incorporates a degree of modularity, which makes
it possible to execute complex computations by dividing them up into a
sequence of ``local'' computations, each involving a subset of the
variables, known as a clique.  This is particularly useful when all
cliques are relatively small.  There exist elegant algorithms to
identify the cliques and streamline the computations, and these have
been implemented in a number of software packages.  In particular, for
a discrete distribution specified by its parent-child conditional
probability tables, the Lauritzen--Spiegelhalter ``probability
propagation'' algorithm \citep{lauritzen1988local} enables efficient computation of all
marginal densities $\{p(x_i): i\in V\}$.  Moreover, for any
$S\subseteq V$ and evidence $X_s = x^*_S$, essentially the same
algorithm computes the prior probability of obtaining that evidence,
$P(X_S = x^*_S)$, and the marginal posterior densities
$\{p(x_i \cd X_S = x^*_S): i\in V\}$.  It is also straightforward to
compute revised probabilities after incorporating external likelihood
evidence relating to a subset of the variables.


%% file: OOBNOverview.tex
A useful extension of the idea of a Bayesian Network is the
Object-Oriented Bayesian Network (OOBN).  This organises the nodes
into a hierarachy of subnetworks, which can greatly simplify
specification and interpretation.  (However, computation still takes
place at the level of the basic nodes).  A particularly valuable
application of OOBNs is based on generic network modules, also termed
classes or fragments, that can be reused, both within and
across higher-level networks \citep{koller2013object}. Here they are 
 described as a ``(network) class''.  Thus a class network  could form part of one or many, more complex networks.
 We indicate such a class by a {\bf boldface} font.  Any specific
	 instantiation of a module in a larger network is set in \texttt{
		 teletype} font (like any other node), while a value (state) of a
 node is indicated by {\em italic\/}.

A class network is like a regular network, except that it can
have interface --- {\em input\/} and {\em output\/} --- nodes, as well
as internal nodes.  Interface nodes are indicated by a grey outer
ring, an input node having a dotted outline, and an output node a
solid outline.  Any network can have nodes that are themselves
instances of other networks, in addition to regular nodes.  Each
instance of a class network within another network is displayed as a
rounded rectangle, which can be expanded if desired to display its
interface nodes.  All instances of a class have identical
probabilistic structure, except that an input node can be
identified, by an incoming arrow, with a node in another network.  A
full description of forensic OOBN networks can be found in
\cite{dawid2007object}.

%% file: CEG_sectionla.tex

The Chain Event Graph (CEG) like the BN is also a diagrammatic representation of
dependence relationships. However, based on an underlying event tree this time
it emphasises the nature of the \emph{stories of the events} that might have
unfolded in time and that provide the different explanations of what has happened. In a legal context these trains of events are often both
a starting point in any analysis of events by police and the end point where
two barristers present to a jury their cases to persuade jurors that the
evidence they present support the story of how things happened to respectively
vindicate or convict a suspect.

In contrast to the BN\ the nodes of the graph of a CEG\ represent the critical
events that might have happened. Its directed edges/arrows then represent the
other events that could have happened in the next step in this unfolding
story. Note here that - unlike for the BN\ - there is no need to first
construct a suitable collection of random variables whose values represent
various instantiations that explain both the defense and the prosecution
cases. Instead the CEG directly represents a natural language description of
the possible stories that explain what happened. It is therefore often easier
to elicit directly from an expert. This is particularly useful in legal cases
where the explanatory trains of events presented by the defence and
prosecution are very different from one another like  when stories are
asymmetrical. 

As for the BN\ the representation has four features: 

\begin{itemize}
\item Temporal sequences: Just as in a verbal explanation the CEG represents a
total order of events (not a partial order as in a BN) consistent with a
particular explanation of how various critical events unfolded.

\item Qualitative: dependence relationships between the unfoldings are represented through its topology and the colouring of its nodes and
edges.

\item Quantitative: through the provision of edge probabilities in its stories,

\item Causal: both directly through the different trains of events it
represents and more formally in terms of the specification of causal algebras
which conjecture what might have happened had someone intervened in the
described processes.  
\end{itemize}

So, like the BN, the CEG provides a formal framework that can be embellished
into a probability model which in the light of a given collection of evidence
supports the calculation of formal measures of support for or against certain
hypotheses. Unlike the BN, instead of representing the model in terms of
relationships between a set of random variables, it represents the case in
terms of its contingent trains of critical events. On the one hand this makes
the topology of the CEG potentially more expressive than the BN. On the other
hand this means that the CEG is usually topologically more complex than its BN
counterpart. 

\subsubsection{Qualitative Structure }

The chain event graph (CEG)\ $G$ has associated with it a coloured directed
acyclic graph (DAG). The finite number of vertices/ nodes of this graph label the
critical events in the stories it represents. The \emph{florets} associated
with each of these nodes are  component subgraphs of depth 1 that describe
the possible next step in the unfolding story leading to an immediately
subsequent next event. A directed edge or arrow connects the root event of the
floret to each of these potential next steps in the story. 

The CEG first takes the event tree of the stories and its florets and colours
two vertices the same if it is believed that the emanating directed edges in
the florets from these two vertices will be the same across the two florets.
Those two vertices are then said to be in the same \emph{stage}. Within this
construction the associated edges are also identified by colours unless this
relationship is immediate from the labelling of the edges within the
represented story. This coloured tree is called a \emph{staged tree}. The CEG
is then constructed by merging those vertices in the staged tree whenever two
vertices have subtrees rooted at those vertices which are topologically and
colour isomorphic. All leaves of the staged tree are then combined into a
single sink vertex. The new vertices of the CEG obtained in this way are
called \emph{positions}. We illustrate this construction process using the very
simple example given below. Other more complex cases, including one describing
a rape case - are given in \cite{Collazzobook} and \cite{Robertson24}. 

The alternative stories told here are of an assailant stabbing a
victim using an identified knife (edge $S_{1}$) or some other knife (edge
$S_{2}$) before discarding or cleaning it and putting it in the cutlery drawer
(respectively edges $D_{i}$ and  $C_{i}$ $i=1,2$). The event tree  depicting
the 4 possible trains of event is given below (on the left). All in court agree that it
doesn't matter what knife was used, the probability of it being placed in the
cutlery drawer rather than discarded would be the same -.i.e. the precise
knife used in the attack is agreed to be independent of where it was
subsequently placed. 

\begin{figure}
\centering
\begin{tabular}{ll}
\includegraphics[scale=0.3, trim={2cm 2cm 4cm 1cm},clip]{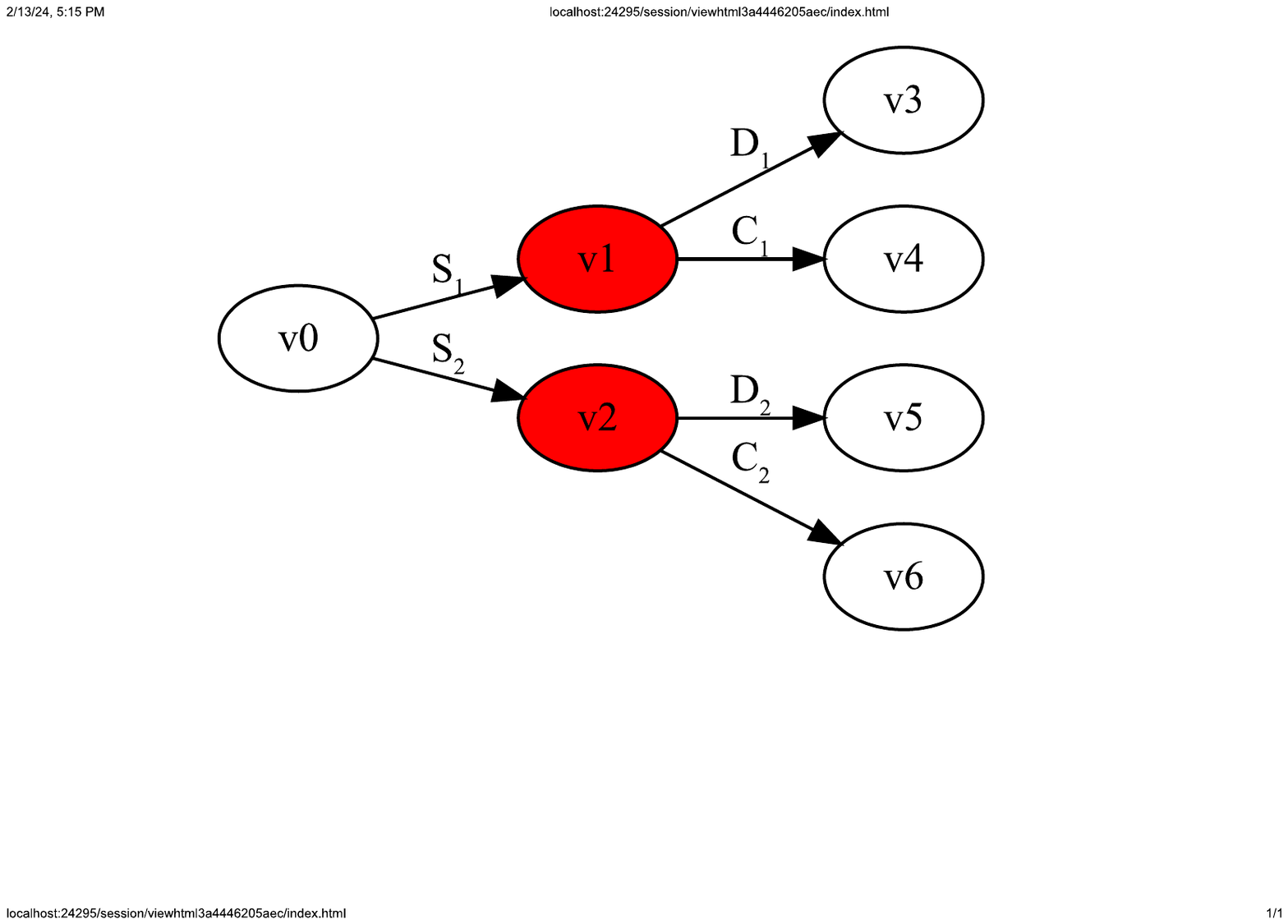}
&
\includegraphics[scale=0.2, trim={1.5cm 2cm 3cm 5cm},clip]{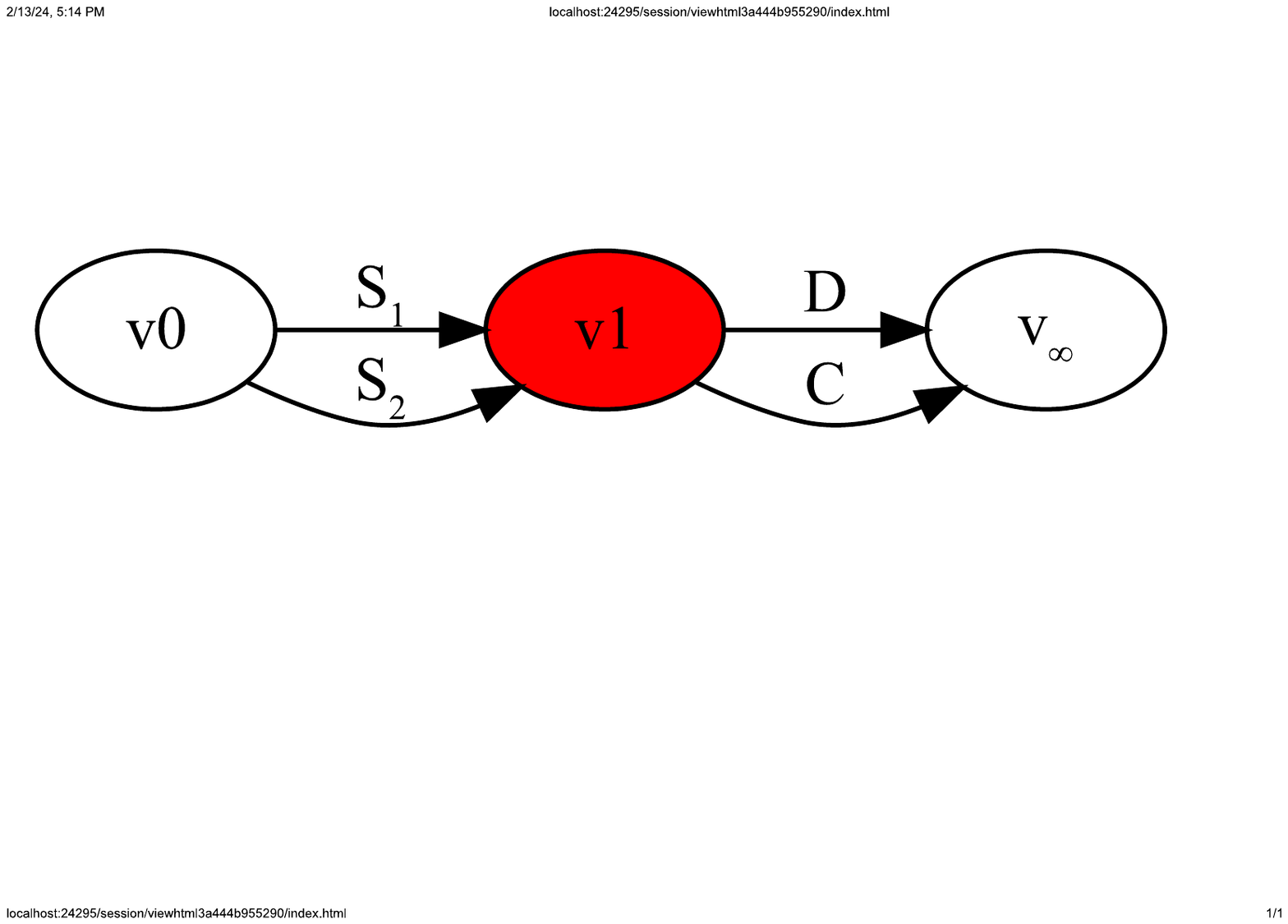}
\end{tabular}
\caption{A staged tree (lhs) and its associated Chain Event Graph (rhs).}
    \label{fig:CEG_example}
\end{figure}

We represent stories associated with these activities below. The  event tree
is transformed into a staged tree by colouring the two vertices at the tail of
$S_{1}$ and $S_{2}$ (here red) and the identified edges the same - here
$D_{1}$ and $D_{2}$ and $C_{1}$ and $C_{2}$. This represents the agreement that the precise knife used in the attack is independent of where it was subsequently placed. Our construction then leads to
the CEG in \figref{CEG_example} (on the right), where nodes $v1$ and $v2$ have been merged. 

%


Note that the CEG now embodies, through its topology, the independence
statement (that whether or not the knife was placed in the cutlery drawer is independent of the precise knife used) whilst retaining an explicit representation of all four possible
unfoldings of events through its root to sink paths. It is easy to prove that
any discrete BN with specified state space has a corresponding CEG
representation where this coloured graph expresses all its implied conditional
independences: see e.g. \cite{SmithAnderson08},\cite{Barclay} and
\cite{Shenvipgm}. But it can also explicitly express through its topology and
colouring a wide range of types of context specific conditional independences,
unlike a BN. It turns out that such structures commonly occur in stories of
crimes. Ways of reading directly implied conditional independences from a CEG
are given in \cite{SmithAnderson08},\cite{Collazzobook} and \cite{Wilkinsonthesis} 

However, the CEG can also represent other types of structural information - not
just irrelevance statements - through its topology: for example that certain
conditional probabilities are logically zero - just by omitting the
corresponding edges in the diagram. See a full discussion of these properties
in \cite{Collazzobook}. 

Once a crime has been committed only certain possible unfoldings of events
will have been possible. For example it will be clear if a victim has died, or it may
have been agreed by all parties before a case comes to court that a particular
victim had been raped and that the contention would focus on who the
perpetrator might be. So although the CEG represents the information we might
have before the presentation of such evidence, after such conditioning on what
is agreed for use in legal cases we present only the root to sink paths that
some parties will argue might describe the course of events after the crime.
By eliminating the paths deemed by all not to provide possible explanations of
what happened  the  CEG can be simplified so that it can represent
those unfoldings that are presented as a possibility by one of the barristers.
This is what is presented here.

\subsubsection{Quantitative Structure and Computation }

Once the CEG has been determined - just as with a BN\ - it can be embellished
into a full probability model. Here we simply add explicit values to the stage
probabilities. The probabilities on the atoms of the event space which are
represented by the root to sink paths of the CEG are calculated simply by
multiplying together the edge probabilities along these paths. Of course some
of these conditional probabilities will be uncertain. However forensic
evidence from sample surveys and experiments can be used to refine these
probabilities using for example the usual Bayesian machinery that calculates
probabilities posterior to such experimental evidence. 

This means, within a legal context, the CEG can be used as a framework for
quantifying uncertainty and in particular from which to construct an
appropriate likelihood ratio. Within a courtroom  it is essential to separate
those probabilities it is legitimate for forensic scientists to provide and
those which should be left to the judgement of judges and/or  jurors. However even when many
such quantifications are the responsibilities of the adjudicators the CEG could
potentially be used to bound probabilities or provide contingent probabilities
for jurors to reflect on: see \cite{Robertson24} and \cite{Collazzobook}  

Computational algorithms for CEGs analogous to BNs have been devised for fast
learning. These are discussed in \cite{ThwaitesCowellsmith},
\cite{Collazzobook} and \cite{Shenvi20theory}. Often, however, the relationship between
forensic data and its source are expressed through their relationships between
continuous observational variables. When this is the case it is convenient to calculate the conditional posterior off-line, possibly using BN software.
The results can then be mapped into the necessary relevant posterior edge probabilities of
the CEG. However it is important to note that - with plausible prior settings
- the modular form of the CEG enables such calculations to take place about
various edge probabilities locally in a justifiable and formal way.

\subsubsection{Object Oriented CEGs}

One disadvantage of using the CEG representation of a case is that when a case
is complex -- like in the Kercher case -- the relevant CEG can be massive.
However like the BN it is modular and so this property can be exploited in the
same way as for a BN. Thus various florets and their probabilistic
embellishments - especially those associated with forensic evidence - 
apply to many cases and can be imported into a case. The modularity also
enables one to zoom down onto parts of the CEG to explore different parts of
the competing stories. Software is currently being developed that would have
such functionality although currently no analogous functionality -- as 
developed for BNs -- is  available.

%% file: BN.tex
We constructed a Bayesian network to represent the logical and
probabilistic relationships between the evidence presented and the
probandum ``Sollecito's knife [\,{\em i.e., exhibit 36}\,] was used to stab
Kercher'', using the proprietary object-oriented Bayesian network
(OOBN) software {\tt Hugin}\footnote{available from {\tt
    https://www.hugin.com}}, version~8.8.

The process of converting the textual description of Section \ref{sec:thelist}
into a Bayesian network involves a number of steps:
\begin{enumerate}
\item Create a node for each item in the list.
\item Create additional nodes, either because they are of independent
  interest, {\em e.g.\/}, the probandum {\tt S knife used?} (was
  Sollecito's knife used to stab Kercher?); or to assist in
  structuring the network, {\em e.g.\/}, {\tt Alternative knife
    (smaller wound)} (the characteristics of an alternative knife
  possibly used on the smaller wound).

  Note that a node represents either an uncertain proposition which
  could be true or false, as in the former case, or a variable with a
  range of possible states, as in the latter case (though we have in
  fact there restricted these to just two: \textit{compatible}, or
 \textit{ incompatible}, with Sollecito's knife).
\item Add appropriate arrows between nodes. 
\end{enumerate}

\begin{sidewaysfigure}[p]
  \centering
  \includegraphics[width=\textheight]{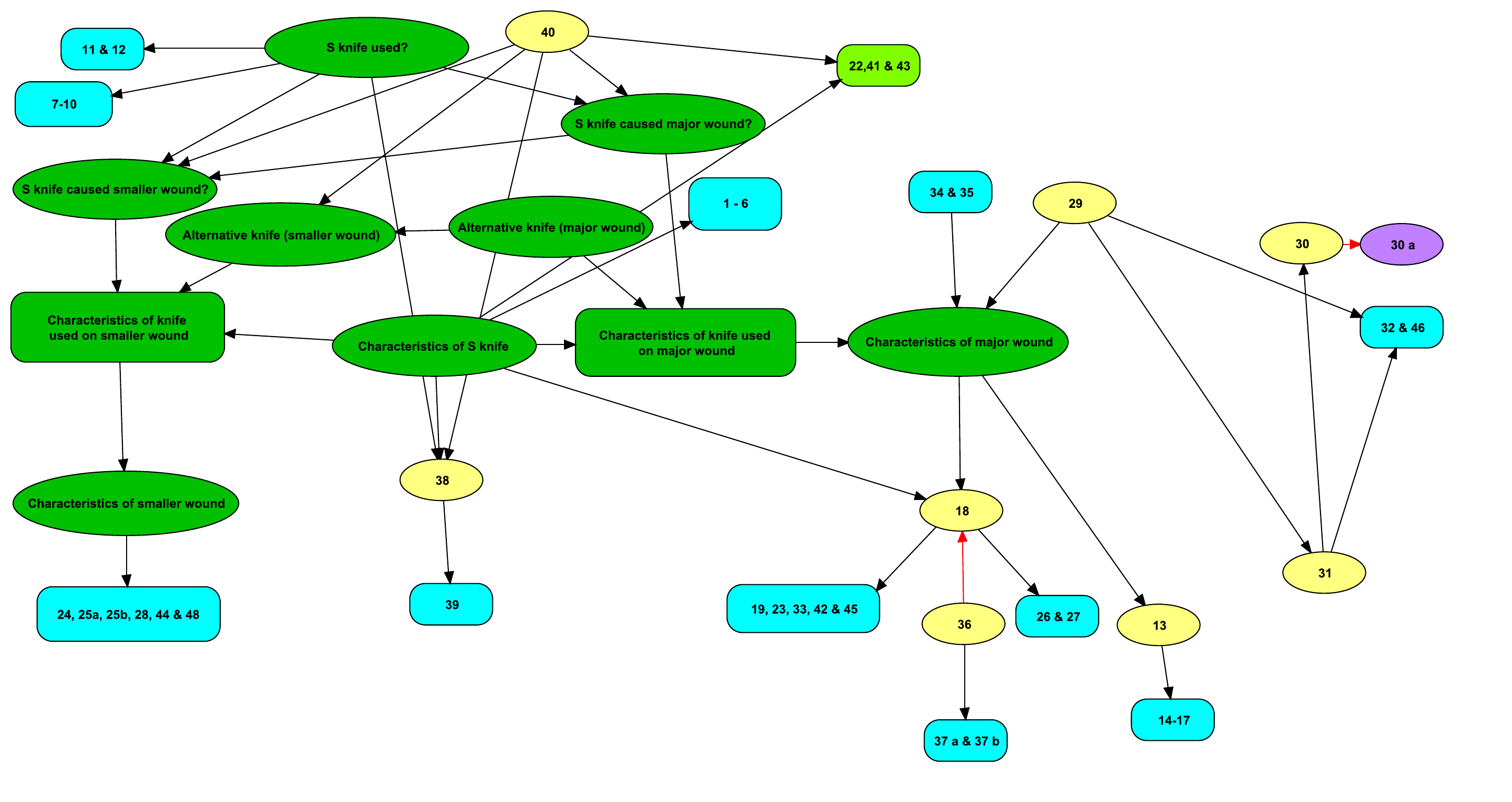}
  \caption{Top-level Bayesian Network}
  \label{fig:bntop}
\end{sidewaysfigure}

\begin{figure}[htbp]
  \centering
  \includegraphics[width=.6\textwidth]{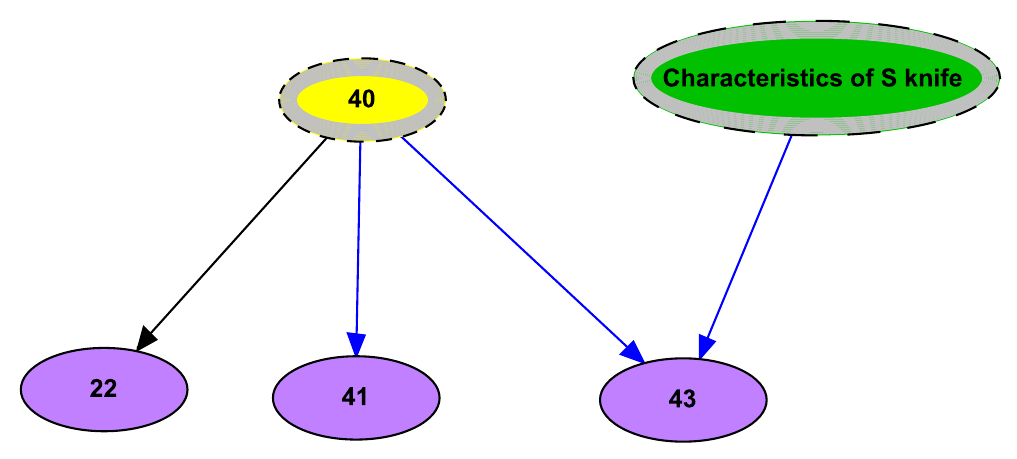}
  \caption{Class network termed {\bf testimony224143} corresponding to the the box labelled {\tt 22, 41 \& 43}  top right corner of the BN shown in \figref{bntop}. This is relative to testimony items 22, 41 and 43 in \secref{thelist}. }
  \label{fig:test22}
\end{figure}

In order to keep the construction and display of the network
manageable, and to assist with interpretation, it has an
object-oriented hierarchical structure, with important propositions
and variables represented at the top level, as displayed in
\figref{bntop}.
Subsidiary variables (in particular, those representing evidence about
propositions) and their relationships are hidden inside subnetworks.
A subnetwork is displayed at the top level by a box with rounded
corners, which can be opened up in the software to reveal its internal
structure.  Thus the box labelled {\tt 22, 41 \& 43} has the internal
structure of the ``class network'' {\bf testimony224143} displayed in
\figref{test22}.  Its nodes {\tt 40} and {\tt Characteristics of S
  knife} are identified with the nodes with the same names at the top
level, while {\tt 22}, {\tt 41} and {\tt 43} represent the evidence
items so numbered in  Section \ref{sec:thelist}.

\begin{figure}[htbp]
  \centering
  \includegraphics[width=.6\textwidth]{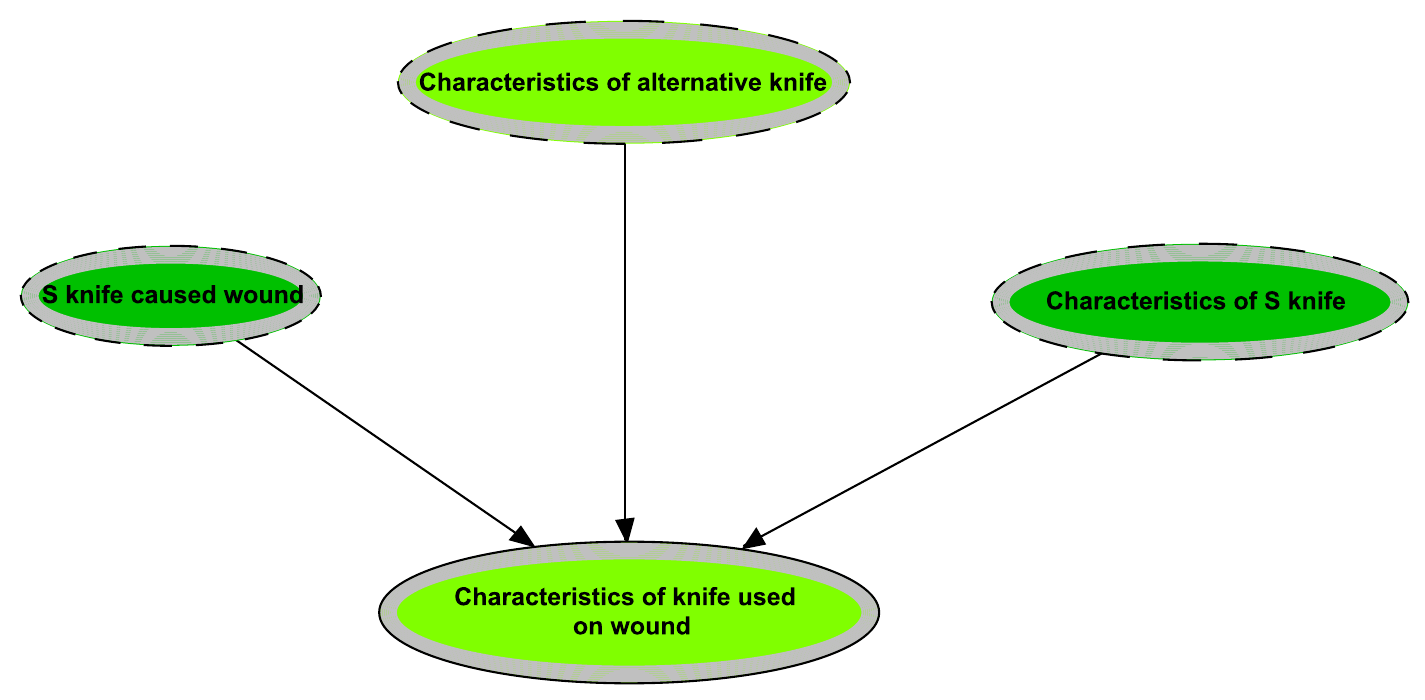}
  \caption{Class {\bf whoseknife}}
  \label{fig:whoseknife}
\end{figure}

Where appropriate, a class network can be reused in distinct places:
thus the class network {\bf whoseknife} displayed in
\figref{whoseknife} abstracts features common to the top-level nodes
{\tt Characteristics of knife used on smaller wound} and {\tt
  Characteristics of knife used on major wound}, these both being
instantiations of that class.

Arrows are initially inserted to represent an intuitive understanding
of (possibly non-deterministic) dependence.  Thus, in
\figref{whoseknife}, {\tt Characteristics of knife used on wound}
depends on whether or not {\tt S knife caused wound?}, on {\tt
  Characteristics of S knife}, and on {\tt Characteristics of
  alternative knife}.  In \figref{test22}, the evidence items {\tt
  22}: ``Liviero could not state whether one or more persons committed
crime'' and {\tt 41}: ``Vinci testimony of compatibility with 40''
({\em i.e.\/}, both wounds could have been made by the same knife)
each depend on {\tt 40}: (whether or not) ``A single knife was used'',
while {\tt 43}: ``Ronchi testimony of compatibility with 40 (and with
exhibit 36)'' further depends on {\tt Characteristics of S knife}.

Some of the relationships represented by arrows are deterministic,
this possibility being aided by judicious choice of variables to
represent.  In \figref{bntop}, {\tt S knife caused smaller wound?}
is fully determined if we know whether or not {\tt S knife used?} (at
all), whether or not {\tt S knife caused major wound?}, and whether or
not (node {\tt 40}), a single knife was used.  If (denoting ``true''
by T, ``false'' by F) the states of these are, in order, TTT, then the
state of {\tt S knife used on smaller wound?} is T; if TTF, then F; if
TFT then F; if TFF then T; while if {\tt S knife used?} is F, then
{\tt S knife used on smaller wound?} is F, no matter the states of the
others.  Similarly, in \figref{whoseknife} the state of {\tt
  Characteristics of knife used on wound} is fully determined by the
states of {\tt S knife caused wound?}, {\tt Characteristics of S
  knife}, and {\tt Characteristics of alternative knife}.

Other relationships are non-deterministic, and have to be described
probabilistically.  This is particularly true of fallible witness
testimony, which is related to, but not necessarily identical with,
the truth of the proposition testified to.  For example, in
\figref{test22}, we might judge that the probability of the testimony
{\tt 41} ``Vinci testimony of compatibility with 40'' is 0.9 when the
proposition {\tt 40} ``A single knife was used'' is true, and 0.2 when
it is false.  We also need to specify the probabilities at founder
nodes, such as the prior probability that {\tt S knife used?}  is
true, or that the characteristics of a possible {\tt Alternative knife
  (major wound)} are compatible with those of Sollecito's knife.
There is clearly a good deal of arbitrariness and subjectivity in such
specifications.

One point to be borne in mind when inserting arrows is that the
absence of an arrow represents (conditional) probabilistic
independence.  Thus in \figref{test22} the lack of any arrows between
nodes {\tt 22}, {\tt 41} and {\tt 43} means that we regard these
testimonies as mutually independent, for any given specific states of
nodes {\tt 40} (``single knife used'') and {\tt Characteristics of S
  knife}.  This might not be reasonable if, for example, the witnesses
had conferred.

 The {\tt Hugin.oobn} files for the complete OOBN together with a \texttt{.html} document containing detailed information about
the network can be found at \url{https://www.dropbox.com/scl/fo/41r3k91q2dzsr4o2ybplb/h?rlkey=35422j6mh7ntg5sc1p0v6747f&dl=0}.

A complete Bayesian network
would specify all the numbers describing just how each variable
depends probabilistically on its predecessors.  We have inserted a few
nominal probability values in the network description, but left most
unspecified.  When all probabilities are specified, one can enter all
the testimony evidence, and then run the software to determine, {\em
  e.g.\/}, the posterior probability of the probandum {\tt S knife
  used?}, in the light of the evidence.


%% file: CEG.tex
\begin{sidewaysfigure}[p]
  \centering
  \includegraphics[width=1.2\textheight, trim={1.2cm 6cm 2cm 1cm},clip]{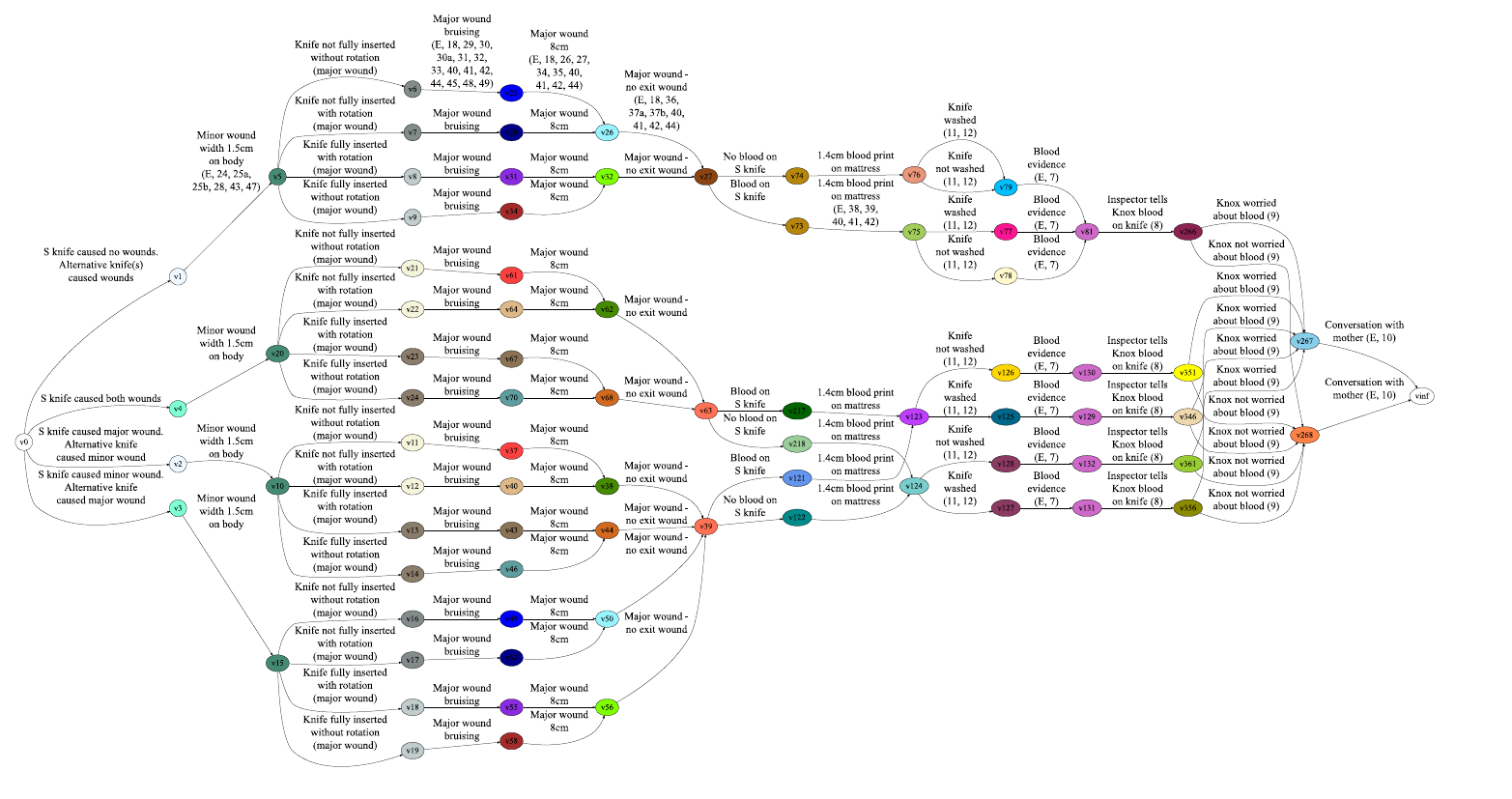}
  \caption{CEG. Relevant items from the list of testimony, propositions and arguments are given in brackets. For readability, where there are lots of repeated edges these are only listed on the top edge.}
  \label{fig:ceg}
\end{sidewaysfigure}

A Chain Event Graph was developed to describe probabilistically the possible ways in which events associated with the use of Sollectio's knife might have unfolded in the case and the evidence associated with these events. This chain event graph is displayed in Figure \ref{fig:ceg}.

Unlike a Bayes Net or Wigmore Chart a CEG can be used to display temporal information, linking the various strands of evidence together as the possible explanations for this evidence unfold.  In the Kercher case we are interested in whether Sollecito's knife was involved in the murder, and specifically whether it was the weapon used for either or both of the knife wounds found on Kercher's body. Thus for this example we set the first four edges in the CEG to be the following four propositions:
\begin{itemize}
\item Sollecito's knife was not used to inflict either of the wounds. An alternative knife or alternative knives were used. 
\item Sollecito's knife was used to inflict both the minor wound and the major wound.
\item Sollecito's knife was used to inflict the major wound. An alternative knife was used for the minor wound.
\item Sollecito's knife was used to inflict the minor wound. An alternative knife was used for the major wound.
\end{itemize}
Conditional on each of these four propositions, we then use the CEG to develop storylines to explain the evidence associated with the knife. Note that as described in Section \ref{sec:wigmore} we do not have access to a word-for-word transcript of the various explanations for the evidence put across by the prosecution and defence so we do not expect this CEG to be a precise representation of the arguments in the case. Nonetheless we can demonstrate how the CEG can be used to aid reasoning and to develop possible explanations for the evidence.

The steps involved in generating the remainder of the CEG were:
\begin{enumerate}
\item We produced a staged tree linking together the evidence and propositions in \ref{sec:thelist} that were relevant to the four main propositions above. Edges associated with evidence (marked with an E in Figure \ref{fig:ceg}) were treated slightly differently to edges associated with propositions in this process. Propositions are uncertain - the aim is to determine how likely the various propositions are given the evidence. For example, we do not know whether the knife was fully inserted into the wound and rotated during the attack, so there are four possible explanations represented in the CEG for this. The evidence in this case is largely agreed on (e.g. the major wound is 8cm) so other possibilities (e.g. the major wound being 7cm) do not need to be represented (the edges can be deleted). The question associated with the evidence edges is what probabilities should be assigned to these edges given the testimony, conditional on the storyline up to that point.
The topology of the staged tree (and hence CEG) can represent the time ordering of events. This is shown for example by the edges describing the blood evidence where a print is left on the mattress, the knife is then washed (or not) and then the blood evidence is taken.  
\item We then coloured the nodes of the staged tree. Nodes with emerging edges that have the same associated probabilities should be the same colour. The probabilities associated with evidence edges can be set with regard to the testimony described in \ref{sec:thelist}. The relevant items in that list are noted in the CEG. For example, the various pieces of testimony relating to whether an exit wound would occur along with the major wound given the different possibilities for which knife was involved (Sollecito's or an alternative) and for how the wound occurred (whether or not it was fully inserted). Some of the edges (e.g. inspector tells Knox blood on knife) are included for clarity of argument and are associated with a probability of one. 


\item We then transformed the staged tree into a chain event graph by merging branches. Two nodes (and the subsequent branches) can be merged if they have the same colour and if the branches to the right hand side of the two nodes are identical (in both structure and colour).
\end{enumerate}

%% file: main.bbl
\begin{thebibliography}{}

\bibitem[Anderson and Twining, 1998]{anderson1998analysis}
Anderson, T. and Twining, W. (1998).
\newblock {\em Analysis of Evidence: How to Do Things with Facts Based on
  Wigmore's Science of Judicial Proof}.
\newblock Law in Context. Northwestern University Press.

\bibitem[Barclay et~al., 2013]{Barclay}
Barclay, L.~M., Hutton, J.~L., and Smith, J.~Q. (2013).
\newblock Refining a bayesian network using a chain event graph.
\newblock {\em International Journal of Approximate Reasoning},
  54(9):1300--1309.

\bibitem[Collazo et~al., 2018]{Collazzobook}
Collazo, R.~A., G{\"o}rgen, C., and Smith, J.~Q. (2018).
\newblock {\em Chain event graphs}.
\newblock CRC Press.

\bibitem[{Corte Assise}, 2009]{sentence}
{Corte Assise} (2009).
\newblock Sentenza di primo grado nel processo contro {Amanda} {Knox} e
  {Raffaele} {Sollecito} per l'omicidio di {Meredith} {Kercher}.
\newblock
  \url{https://www.giurisprudenzapenale.com/wp-content/uploads/2023/11/ASSISE-PERUGIA-KNOX-SOLLECITO.pdf}.
\newblock G. Massei Presidente est., B. Cristiani, Giudice Est.

\bibitem[Dawid, 1979]{dawid1979conditional}
Dawid, A.~P. (1979).
\newblock Conditional independence in statistical theory.
\newblock {\em Journal of the Royal Statistical Society Series B: Statistical
  Methodology}, 41(1):1--15.

\bibitem[Dawid et~al., 2007]{dawid2007object}
Dawid, A.~P., Mortera, J., and Vicard, P. (2007).
\newblock Object-oriented {B}ayesian networks for complex forensic {DNA}
  profiling problems.
\newblock {\em Forensic Science International}, 169(2-3):195--205.

\bibitem[Dawid et~al., 2011]{Dawid2011-DAWIN}
Dawid, P., Schum, D., and Hepler, A. (2011).
\newblock Inference networks: {B}ayes and {W}igmore.
\newblock In Dawid, P., Twining, W., and Vasilaki, M., editors, {\em Evidence,
  Inference and Enquiry}, page 119. Oup/British Academy.

\bibitem[Gill, 2014]{gill2014misleading}
Gill, P. (2014).
\newblock {\em Misleading DNA Evidence: Reasons for Miscarriages of Justice}.
\newblock Academic Press. Elsevier Science \& Technology Books.

\bibitem[Gill, 2016]{GillKercher}
Gill, P. (2016).
\newblock Analysis and implications of the miscarriages of justice of {A}manda
  {K}nox and {R}affaele {S}ollecito.
\newblock {\em Forensic Science International: Genetics}, 23:9--18.

\bibitem[Kadane and Schum, 2011]{kadane2011probabilistic}
Kadane, J.~B. and Schum, D.~A. (2011).
\newblock {\em A probabilistic analysis of the Sacco and Vanzetti evidence}.
\newblock John Wiley \& Sons.

\bibitem[Koller and Pfeffer, 2013]{koller2013object}
Koller, D. and Pfeffer, A. (2013).
\newblock Object-oriented {B}ayesian networks.
\newblock {\em arXiv preprint arXiv:1302.1554}.

\bibitem[Lauritzen and Spiegelhalter, 1988]{lauritzen1988local}
Lauritzen, S.~L. and Spiegelhalter, D.~J. (1988).
\newblock Local computations with probabilities on graphical structures and
  their application to expert systems.
\newblock {\em Journal of the Royal Statistical Society: Series B
  (Methodological)}, 50(2):157--194.

\bibitem[Robertson et~al., 2024]{Robertson24}
Robertson, G., Wilson, A., and Smith, J.~Q. (2024).
\newblock Chain event graphs for assessing activity level propositions in
  forensic sccience in relation to drug traces on bank notes.
\newblock In preparation.

\bibitem[Shenvi and Smith, 2020a]{Shenvipgm}
Shenvi, A. and Smith, J.~Q. (2020a).
\newblock Constructing a chain event graph from a staged tree.
\newblock In {\em International Conference on Probabilistic Graphical Models},
  pages 437--448. PMLR.

\bibitem[Shenvi and Smith, 2020b]{Shenvi20theory}
Shenvi, A. and Smith, J.~Q. (2020b).
\newblock Propagation for dynamic continuous time chain event graphs.
\newblock https://arxiv.org/abs/2006.15865v1.

\bibitem[Smith and Anderson, 2008]{SmithAnderson08}
Smith, J.~Q. and Anderson, P.~E. (2008).
\newblock Conditional independence and chain event graphs.
\newblock {\em Artificial Intelligence}, 172(1):42--68.

\bibitem[Thwaites et~al., 2008]{ThwaitesCowellsmith}
Thwaites, P., Smith, J.~Q., and Cowell, R.~G. (2008).
\newblock Propagation using chain event graphs.
\newblock {\em Uncertainty in Artificial Intelligence}, pages 546--553.

\bibitem[Wilkerson, 2020]{Wilkinsonthesis}
Wilkerson, R. (2020).
\newblock {\em Bayesian Graphcial Models and Chain Event Graphs}.
\newblock Ph{D} thesis, University of of Warwick.

\end{thebibliography}
